# 2D beam shaping via 1D spatial light modulation

James E. M. Whitehead,[1] Albert Ryou,[1] Shane Colburn,[1] Maksym Zhelyeznyakov,[1] and Arka Majumdar[1,2,*]

[1]Department of Electrical and Computer Engineering, University of Washington, Seattle, Washington 98195, USA.
[2] Department of Physics, University of Washington, Seattle, Washington 98195, USA.
*arka@uw.edu

Many emerging reconfigurable optical systems are limited by routing complexity when producing dynamic, two-dimensional (2D) electric fields. Using a gradient-based inverse designed, static phase-mask doublet, we propose an optical system to produce 2D intensity wavefronts using a one-dimensional (1D) intensity Spatial Light Modulator (SLM). We show the capability of mapping each point in a 49 element 1D array to a distinct $7 \times 7$ 2D spatial distribution. Our proposed method will significantly relax the routing complexity of 2D sub-wavelength SLMs, possibly enabling next-generation SLMs to leverage novel pixel architectures and new materials.

Fast, dynamic manipulation of two-dimensional (2D) optical fields is integral to many emerging applications including optical holography [1], non-line-of-sight imaging [2], optical neural networks [3] and imaging through disorder [4]. Currently, most of these applications rely on either digital micromirror devices (DMDs) or liquid crystal-based spatial light modulators (SLMs). Both technologies suffer from low-speed operation (~1-100 kHz). Several phase shift mechanisms, such as electro-optic modulation via free carrier dispersion [5] or the Pockels effect [6], can potentially increase the speed by several orders of magnitude. An increase in operating speed, however, must be accompanied by a reduction in switching energy per pixel to maintain an acceptable level of power consumption. Reducing this switching energy requires $V_m < \lambda^3$, where $V_m$ is the active pixel mode volume and $\lambda$ is the operating wavelength and is largely independent of the reconfiguration mechanism being used [7]. Unfortunately, accommodating a high pixel count 2D SLM with sub-wavelength pitch poses a significant routing challenge. A 2D SLM with $N \times N$ pixels will need $N^2$ control signals with an electrode spacing of $\sim\lambda/N$ to address all pixels along each row. At visible wavelengths, even with the most sophisticated fabrication techniques, $N$ will be limited to ~10 without using vias, which also do not scale. The number of control signals can be reduced to $O(N)$ by exploiting active electronics, including a thin film transistor-based active matrix. While such active electronics are a norm in displays, realizing full display high-speed operation (~100 MHz) with active matrix over a reasonable number of pixels remains difficult [8].

A 1D-SLM, however, is far simpler to address and the routing of the electrical control lines becomes trivial. There are several recent demonstrations of 1D-SLM exploiting free-carrier dispersion effects [9]. Unfortunately, 1D beam shaping is not as versatile as 2D wavefront modulation. An attractive solution is to map a 1D array of $N^2$ phase-shifters to a 2D array of $N \times N$ pixels, which can perform the 2D optical wavefront modulation without difficulty in routing the electrical control signals. To support this, a recent work [10] used a random medium to enable a 1D-to-2D mapping for imaging. Here a high speed (350kHz) 1D SLM is used to focus light in a 2D plane through the random medium. However, arbitrary wavefront shaping, as needed for a 2D SLM, was not demonstrated.

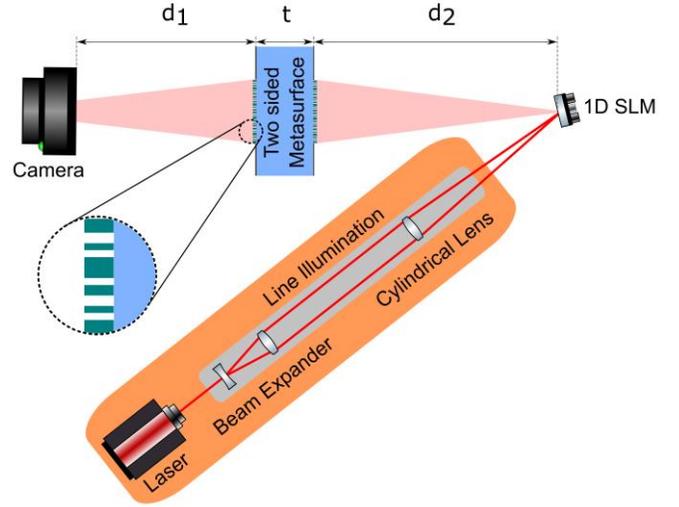

Figure 1. Proposed optical architecture for 1D-to-2D transformer. Coherent light is focused into a line to efficiently illuminate a linear array of 1D SLM pixels. Light is sent to a pair of meta-optical structures and into the camera, creating arbitrary 2D patterns. The gap between the 1D SLM and the first phase mask is kept at $d_2 = 2\ mm$. The separation between masks is $t = 1\ mm$, corresponding to a standard glass wafer thickness so that masks can be fabricated on either sides of a substrate [11]. The distance between the final mask and the camera is $d_1 = 2cm$.

In this paper, we propose a method to control a 2D wavefront by varying the pixels arranged in a 1D array. The key is an inverse-designed multi-layer phase mask that maps an input point source to a two-dimensional spatial field profile that forms an orthogonal 2D intensity basis. To demonstrate this, we optimize a 1D-to-2D transform that maps pixels from a 1D SLM to an effective 2D SLM. Figure 1 shows the proposed optical architecture for the 1D-to-2D SLM. The laser light is modulated using a 1D array of tunable pixels, which then passes through the composite phase masks. We design these two phase-masks to route the light from each 1D pixel to produce a desired 2D intensity distribution at a specific plane.

We first construct a forward model that simulates the light propagation from a 1D SLM through free space and include the light's interaction with discretized phase masks. We employ the band-limited angular spectrum method to simulate the forward propagation [11]. The light with, a wavelength $\lambda$, from the input source $E(x, y)$ is propgataed through a distance of $d$ in free space.

$$E(x, y, z_0) = E(x, y, 0) * h(x, y, z_o)$$
$$\mathcal{F}\{h(x, y, z_o)\} = e^{j\frac{2\pi}{\lambda}z_0\sqrt{1-(\lambda f_x)^2-(\lambda f_y)^2}} \times A(f_x, f_y)$$

$E(x, y, z)$ is the electric phasor field, $h(x, y, z)$ is the point spread function, $\mathcal{F}$ is the 2D spatial Fourier transform operator, and $z_0$ is the propagation distance. $A(f_x, f_y)$ is a mask that limits the spatial bandwidth to be lower than $f_x$ and $f_y$, which blocks high-angle wavevector components that would otherwise wrap around and re-enter the simulation domain [11]. The effect of the phase-mask is modeled by a point-by-point multiplication of the input field with the phase mask's complex amplitudes.

This forward model is then used in an automatic differentiation-based optimization method to design the phase profiles which will map the 1D pixel array to a 2D wavefront. For the optimization, we construct a cost function $C$ based on the desired input-output mapping

$$C = -\prod_k^N \left( \sum_{i,j} y_k^{(i,j)} \hat{y}_k^{(i,j)} \right)$$

where $y_k^{(i,j)}$ and $\hat{y}_k^{(i,j)}$ are the target and the output spatial distributions respectively for the $k^{th}$ input mode (in this case each pixel in the 1D array). $N$ is the number of pixels in the 1D array. The two-dimensional field is discretized in the (x, y) plane with $i$ and $j$ being the discretization indices. This cost function is designed to make the output modes similar to the desired modes while ensuring that each output mode contains similar power. We note that binning power in the target output pixel outperforms using a mean squared error cost function since the output pixel amplitudes do not need to be constrained.

The phase masks are iteratively updated to minimize the cost function until the system converges. Gradient-based optimization is used to adjust the phase based on the cost function. The gradient, with respect to all pixels, is calculated using auto-differentiation facilitated by the graph-based linear algebra library TensorFlow [12]. The phases are updated using the ADAM optimizer implementation in TensorFlow.

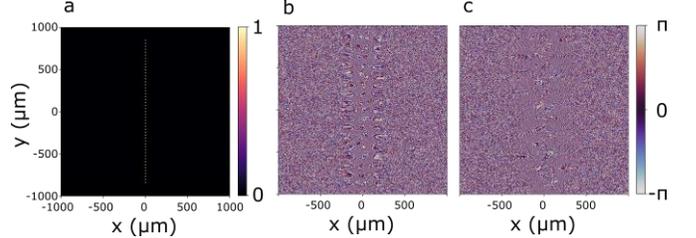

Figure 2. a) Input pixel intensity, optimized phase profiles for the a) first and b) second phase mask.

Figure 2b and 2c show optimized phase-masks for 1D-to-2D mapping. We impose a square aperture for the phase masks with dimension of $2\ mm \times 2\ mm$. The 1D SLM pixel pitch is $p = 25\ \mu m$.

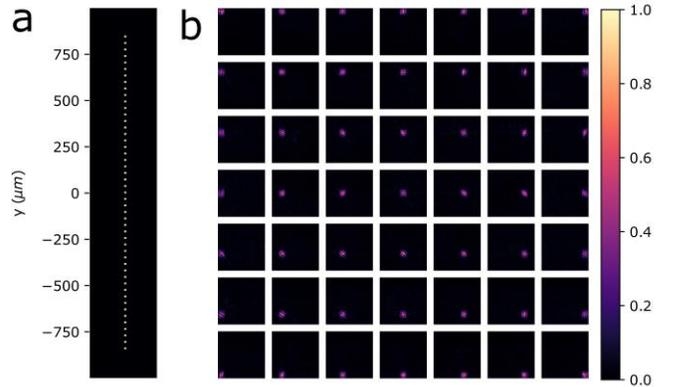

Figure 3. a) Input 49 element 1D SLM with all input pixels illuminated b) Simulated output modes for 1D 49 element SLM input.

Using these phase masks, we can map a 1D array to a 2D field distribution. When excited by individual pixels from the 1D SLM, the output field at the 2D SLM output will be illuminated. The output fields are well defined within the output pixel boundary.

Figure 3a shows the 1D array of 49 spots. Each spot maps to a specific point in the 2D plane, as shown in the $7 \times 7$ array in figure 3b, when passing through two phase-masks. The 49 points in the resulting 2D array will approximate a complete intensity basis in the $7 \times 7$ output space. We note that the choice of 49 pixels is limited by the available computational resources. More pixels can be added to increase the image resolution. However, the pitch of the input 1D

SLM is limited due to diffraction, as explained later in the paper.

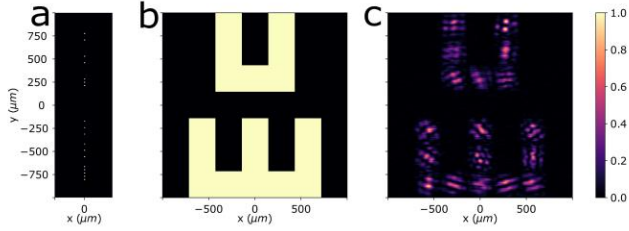

Figure 4. a) 1D SLM input field b) Target output field c) Output field

By choosing which input pixels are illuminated, 2D patterns can be generated. Figure 4 shows the projection of an arbitrary pattern: the letters "UW".

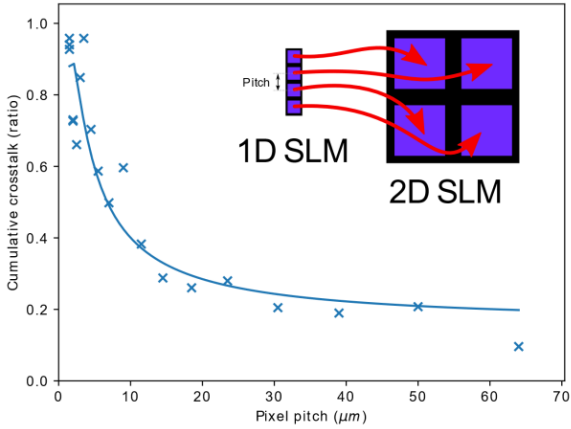

Figure 5. Cumulative crosstalk between output modes for a range of trained systems that map a 4-pixel 1D SLM to a 2 × 2 output SLM. The simulation region was shrunk by a factor of 10 to clearly observe the breakdown of the mapping. All other simulation parameters are identical to the featured model except for the swept input pixel pitch. The solid curve serves as a guide to the eye.

Finally, we analyze how diffraction poses a limit on the proposed 1D-to-2D transformation. Light from different pixels in the 1D array impinge on the phase masks at different angles, and thus the reshaping of the 1D array to a 2D space depends on the angular resolution of the phase-mask doublet. For the arrangements of the optics simulated earlier, the Abbe diffraction limit (presuming a circular aperture) would be 633 nm. We hypothesize that when the pixel pitch approaches this limit, our 1D-to-2D mapping fails. To validate this hypothesis, we reduce the sizes of the phase masks by a factor of 10 (in order to reduce the pixel pitch while keeping memory use below the limit) and sweep the input pixel pitch. We optimize a set of otherwise identical systems that map a 4 pixel, 1D SLM to a 2 × 2 pixel output and quantify the error as the cumulative crosstalk (Figure 5). A crosstalk of 1 indicates that energy from a single input pixel is distributed across all output pixels while a crosstalk of 0 means that all power is in the intended output. This value is summed across all the inputs for a given simulation to give the cumulative crosstalk.

In this reduced simulation, the calculated Abbe diffraction limit is $6.33 \, \mu m$. As expected, the performance of the phase mask doublet markedly decreases as the input pixel pitch approaches the diffraction limit.

The proposed method can be experimentally verified with commercial SLMs. The dimensions used in our simulations are motivated by experimental feasibility, such as the dimensions of the beam expanders and spacing between optics. The phase masks can be implemented using metasurfaces [13]. Metasurfaces are subwavelength diffractive optics that can shape the phase of incident light with high spatial resolution [14,15]. Under a local phase approximation, metasurfaces can be modelled as a phase mask [16]–[18]. The phasor response of the phase mask elements can be simulated using a rigorous full wave electromagnetic simulation (finite-difference time-domain or rigorous coupled-wave analysis [14,19]). Finally, individual optical scatterers would be mapped to best approximate the phase profiles yielding functional meta-optical structures. The doublet can be realized by fabricating two metasurfaces on both sides of a glass slide. Additionally, mm-aperture, visible wavelength metasurfaces have already been reported and can potentially be scaled to even larger apertures.

In summary, we demonstrate a method to generate 2D varying intensity profiles from a 1D SLM using a pair of inverse designed phase masks. A reshaping operation is performed to illustrate the effectiveness of the system. Finally, we propose a way to experimentally realize this system which can be used to make novel high-speed 2D SLMs due to reduced electronic routing complexity.

All the simulations were performed on a Nvidia 1070Ti 8GB, Intel i5-4690 CPU 3.5 GHz 4-Core, 8GB System Memory.

**Funding.** The research is supported by a Samsung GRO grant and NSF-DMR-2003509.

**Acknowledgements.** We acknowledge useful discussion with Prof. Rafael Piestun.

**Disclosures.** The authors declare no conflicts of interest.